\documentclass[sigconf]{acmart} 

\usepackage{tabu} 
\AtBeginDocument{%
  }

\acmConference[Conference acronym AHs'25]{Make sure to enter the correct
  conference title from your rights confirmation emai}{March 16--20,
  2025}{Abu Dhabi, UAE}
\acmISBN{978-1-4503-XXXX-X/18/06}

\begin{document}

\title{Happiness Finder: Exploring the Role of AI in Enhancing Well-Being During Four-Leaf Clover Searches}

\author{Anna Yokokubo}
\authornote{Both authors contributed equally to this research.}
\email{anna.yokokubo@koshizuka-lab.org}
\orcid{0000-0003-2657-4961}
\author{Takeo Hamada}
\authornotemark[1]
\email{takeo.hamada@koshizuka-lab.org}
\affiliation{%
  \institution{The University of Tokyo}
  \city{Bunkyo-ku}
  \state{Tokyo}
  \country{Japan}
}

\author{Tatsuya Ishizuka}
\affiliation{%
  \institution{The University of Tokyo}
  \city{Bunkyo-ku}
  \country{Japan}}

\author{Hiroaki Mori}
\affiliation{%
  \institution{The University of Tokyo}
  \city{Bunkyo-ku}
  \country{Japan}
}

\author{Noboru Koshizuka}
\affiliation{%
 \institution{The University of Tokyo}
 \city{Bunkyo-ku}
 \state{Tokyo}
 \country{Japan}
 }
\email{noboru@koshizuka-lab.org}

\renewcommand{\shortauthors}{Yokokubo and Hamada et al.}

\begin{abstract}
A four-leaf clover (FLC) symbolizes luck and happiness worldwide, but it is hard to distinguish it from the common three-leaf clover. While AI technology can assist in searching for FLC, it may not replicate the traditional search's sense of achievement. This study explores searcher feelings when AI aids the FLC search. 
In this study, we developed a system called ``Happiness Finder'' that uses object detection algorithms on smartphones or tablets to support the search. 
We exhibited HappinessFinder at an international workshop, allowing participants to experience four-leaf clover searching using potted artificial clovers and the HappinessFinder app. This paper reports the findings from this demonstration.

\end{abstract}


\begin{teaserfigure}
  \centering
  \includegraphics[width=\textwidth]{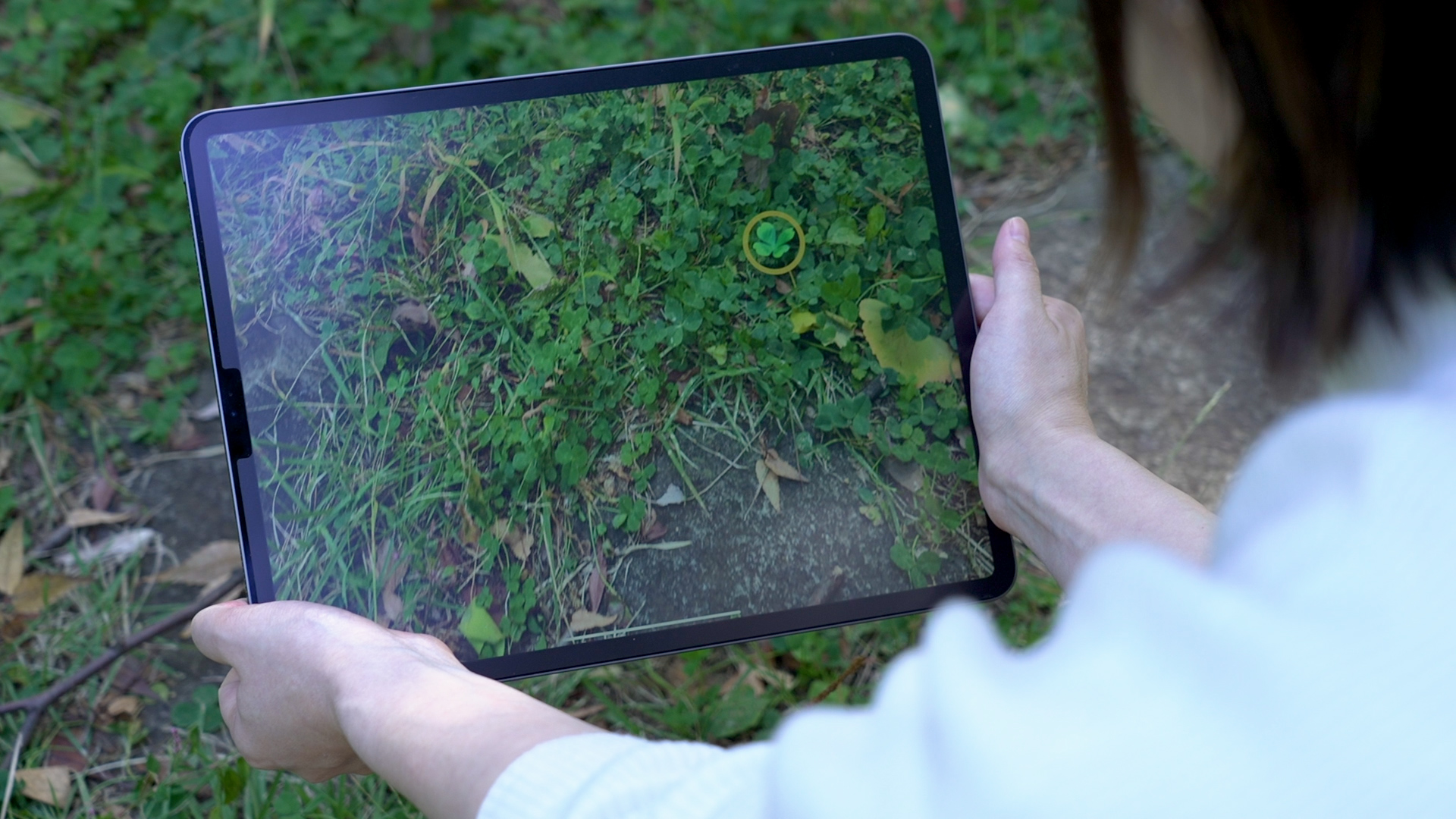}
  \caption{Users can take a picture of clovers with a mobile device's camera to know if an FLC is in the captured image with object detection techniques. However, would they be happy to find the FLC through such technological intervention?}
  \label{fig:teaser}
\end{teaserfigure}


\maketitle

\section{Introduction}

Happy people tend to be healthier and have better relationships \cite{lyubomirsky2005benefits}. Lyubomirsky et al. propose that happiness is determined by genetic factors, environmental influences, and personal activities \cite{lyubomirsky2005pursuing}. While the first two are difficult to control, happiness can be influenced by one's own actions.
Happiness is based on short-term positive emotions, while well-being reflects a broader, desired state in life. Positive psychology promotes well-being, emphasizing proactive engagement. Seligman’s well-being theory (PERMA\cite{seligman2011flourish}) highlights key elements such as positive emotions and meaning
. However, increasing reliance on advanced technology may diminish human agency.
A four-leaf clover (FLC) is considered a symbol of luck, occurring in about 1 in 10,000 clovers \cite{DavidBradley2008}. Aristotle described luck as irregular and indeterminate \cite{waterfield1996physics}. Some believe that taking an FLC found by someone else brings bad luck \cite{beckwith1923signs}.

In this paper, we aim to explore how searchers feel when technology assists them in their search for FLC. To this end, we have developed {\sl Happiness Finder}, a system that supports the search for FLC using an object detection algorithm. We employed camera images from smartphones and tablets, which are widely used by the general public, to avoid the novelty effect caused by state-of-the-art devices that overly stimulate the user's interest. 

\section{Related Work}
Digital technology has rapidly evolved around efficiency, speed, and instantaneity, pressuring individuals to remain constantly connected and produce more in less time. This "always-on" culture can lead to anxiety and reduced mental space. While concepts like "Digital Detox" \cite{10.1145/3334480.3382810}\cite{10.1145/3544549.3585681} and "Balanced Digital Lifestyles" (e.g., Hygge \cite{Wiking_Hygge2016}, Ikigai \cite{Garcia_Ikigai2017}) aim to reduce stress by disconnecting from technology, such approaches are often impractical in urban environments. Instead, we propose ``Slow Digital,'' which promotes mindful technology use to enhance well-being rather than rejecting digital conveniences. This aligns with positive computing by fostering meaningful engagement and reducing anxiety, shifting HCI from passive consumption to more balanced and sustainable digital experiences.

Human object searching generally serves two purposes: finding an item quickly, as in locating lost keys or defective parts, and enjoying the search process itself, even when success is uncertain. The first category has been well explored with digital tools such as RFID \cite{Sasagawa:2016, Komatsuzaki:2011}, trackers (e.g., AirTag), cameras \cite{Yan_CamFi:2022, Latina:2023}, and GPS \cite{Pachipala:2022}. The second category includes searches for hidden elements like "Where’s Wally?" or "Hidden Mickey," as well as symbols of happiness, such as shooting stars or four-leaf clovers (FLC). While some studies focus on FLC datasets \cite{BPPA_FGVC6} or HMD-based searches \cite{hamada2020finding}, the impact of AI-assisted FLC discovery on well-being remains unclear. Unlike functional searching, there is still room to explore digital tools that enhance the emotional value of the search process and the rarity of the target itself.

\section{Happiness Finder}
\begin{figure}[t]
  \centering
  \includegraphics[width=0.45\textwidth]{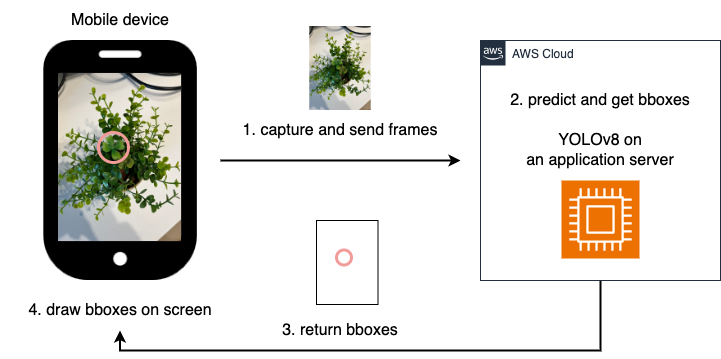}
  \caption{Overview of the Happiness Finder system. The camera footage is captured on a mobile device and sent to a backend server in the cloud. The object detection model is executed on the server, and the resulting BBoxes are returned to the mobile device.}
  \label{fig:system_overview}
\end{figure}

To assist in the search for FLC, we developed {\sl Happiness Finder}, a web application using object detection techniques (Fig. \ref{fig:system_overview}). A real-time object detection model (Ultralytics YOLOv8x~\cite{yolov8_ultralytics}) was trained on 1234 images of an FLC dataset~\cite{BPPA_FGVC6}. These images were annotated using an AI-assisted annotation tool (roboflow \cite{roboflow}). The object detection model infers camera footage acquired every second from a smartphone, whether or not FLC appears in the image, and if so, the coordinate of the Bounding Box (BBox) representing FLC is obtained. 
The precision, recall, and mean Average Precision (mAP@0.5) metrics were 0.676, 0.500, and 0.533, respectively, with 310 dataset images. 
Considering the image file size (approximately 500 KB) and mobile connection uplink speed (1 Mbps), we set the one-second intervals for object detection in a stable communication environment.
When the BBox scores higher than the specific confidence score set by the user, a sound \footnote{\url{https://soundeffect-lab.info/sound/button/mp3/decision3.mp3}} is played, and the BBox is superimposed on the camera image displayed on the web browser 
.

 \section{Study}
We conducted an exhibition and preliminary experiment of Happiness Finder at an international conference in Spain. 
An experimenter explained the overview and usage of Happiness Finder to the participants for approximately three minutes. 
Following this, participants began searching for FLC using either their own smartphones or a tablet device provided by the experimenter. 
After discovering FLC through Happiness Finder, participants were asked to respond to a survey about their experience with Happiness Finder. 
The experimenter placed pots containing artificial FLC on and around the table.

Twenty-two participants took part in the experiment, with the following demographic breakdown: five from Asia (two males, three females), 12 from Europe (nine males, three females), two from the Middle East (one male, one female), one from Central and South America (male), and two from North America (two females). 
The average age range of all participants was 30-39 years.


We consisted of the following nine questions.
All responses to the questions were in a multiple-choice format, and the answers to questions (5) and (6)were based on a 7-point Likert scale.

\begin{enumerate}
    \item What country are you from?
    \item What is your age group?
    \item What is your gender?
    \item Did you know that a four-leaf clover is considered a symbol of happiness? Is there such a cultural belief in your country?
    \item I would be happy to find a four-leaf clover.
    \item How happy were you if you found a four-leaf clover with "Happiness Finder" compared to if you found it usually without a smartphone?
    \item How did you find a four-leaf clover?
    \item Who did you feel found the four-leaf clover?
    \item Feel free to write any additional comments here.
\end{enumerate}

\section{Results}

\begin{figure}[t]
  \centering
  \includegraphics[width=0.45\textwidth]{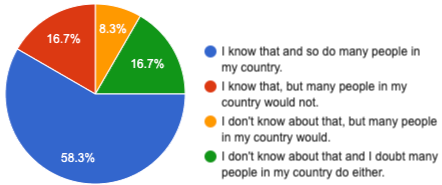}
  \caption{Results of responses to survey Question (4).}
  \label{fig:ps_q4ans}
\end{figure}

The nine-item survey results are analyzed per question. Conducted at an international exposition, this study collected responses from diverse regions, including Asia, Europe, the Middle East, and the Americas.
For question (4), 75\% of participants recognized FLC as a symbol of good luck (Fig. \ref{fig:ps_q4ans}), though awareness was lower in parts of the Middle East and Asia.
For questions (5) and (6). 90.9\% of participants felt happy finding an FLC, while only 4.5\% responded negatively. Question (6) compared experiences of finding FLC naturally versus using Happiness Finder. 68.1\% had a positive impression of the latter, confirming its favorable reception.
For question (7), 87.5\% believed Happiness Finder located the FLC, while 25\% saw it themselves first. In question (8), 66.7\% credited the system, and 12.5\% found it independently.
Open-ended feedback was limited, but some participants commented:``Happiness Finder is a cute system.,'' ``Happy to have found a four-leaf clover.'' and ``I’d like to find hard-to-spot objects, like insects, using it.''
These results suggest that Happiness Finder enhances the joy of searching for FLCs.

\section{Conclusion and Future Work}

We developed Happiness Finder, a tool to support the search for FLC using smartphones or tablets, and examined how the experience of searching for FLC with AI affects subjective well-being. 
Through our study, we observed a tendency for many participants to feel happiness upon discovering a four-leaf clover and to have a positive impression of the experience using Happiness Finder.

From a future perspective, 
we aim to implement additional features to serve as a search support partner without losing user agency and to design and model support levels for an active searching experience.


\begin{acks}
This research was supported by the H-UTokyo Lab.

\end{acks}

\bibliographystyle{ACM-Reference-Format}
\bibliography{sample-base}

\appendix









\end{document}